\documentclass[twocolumn,showpacs,preprintnumbers,amsmath,amssymb]{revtex4}

\usepackage{graphicx}
\usepackage{dcolumn}
\usepackage{bm}

\begin{document}

\preprint{APS/123-QED}

\title{Even-Odd Oscillation in Conductance of Single-Row Sodium Nanowire}

\author{Yoshiyuki Egami}
\affiliation{\mbox{Graduate School of Engineering, Osaka University, Suita, Osaka 565-0871, Japan.}}

\author{Tomoya Ono}
\affiliation{\mbox{Graduate School of Engineering, Osaka University, Suita, Osaka 565-0871, Japan.}}

\author{Kikuji Hirose}
\affiliation{\mbox{Graduate School of Engineering, Osaka University, Suita, Osaka 565-0871, Japan.}}

\date{\today}

\begin{abstract}
We present a first-principles calculation of the electronic conduction properties of single-row sodium nanowires suspended between semi-infinite electrodes. The conductance of the nanowire is $\sim 1\ {\rm G_0}$ ($=2e^2/h$) and oscillates with a two-atom period as the number of the atoms within the nanowire varies. Moreover, we observed bunches of high electron density with a two atom-lengths in the channel density distribution. The relation between the period of the conductance oscillation and the length of bunches are examined by using simplified models and is found to be largely affected by the characteristics of the infinite wire.
\end{abstract}

\pacs{73.63.Nm, 68.65.La, 73.21.Hb, 73.23.Ad}
\maketitle


The properties of condensed systems with nanometer dimensions have attracted considerable interest in the past decade because of their great importance in fundamental physics as well as in nanometer-scale fabrication technologies. When the systems' dimensions reach the order of the electron mean free path, the electron transport becomes ballistic and its conductance is quantized in the unit of ${\rm G_0}$ ($=2e^2/h$) contrary to the diffusive transport in macroscopic systems, where $e$ is the electron charge and $h$ is Planck's constant. Since clearly showing such unique quantum phenomena, the metallic nanowire contacts have been examined in many experimental and theoretical studies.
In experiments, metallic nanowires are generated using scanning tunneling microscopy or mechanically controllable break junctions\cite{datta-ruiten}. Many experiments observed that the nanowires made of monovalent atoms such as Na\cite{krans1,krans2}, Cu\cite{krans2,hansen,costa1}, Ag\cite{hansen,costa2} or Au\cite{hansen,costa1,ohnishi,ludoph} manifest conductance quantization. Moreover, Smit {\it et al.}\cite{smit} found in their recent experiment that the conductances of Au, Pt and Ir nanowires oscillate with a two-atom period as the number of atoms constituting the nanowire varies, which is the so-called even-odd oscillation.
On the theoretical side, there have been several first-principles studies on the electron conduction properties of single-row nanowires, and some of them reported that the conductances of the single-row nanowires suspended between electrodes exhibit the even-odd oscillation\cite{sim,lang1,lang2,lang3,emberly,kobayashi1,kobayashi2,tsukamoto,lee,egami,petr}. Although these results are consistent with those of previous experiments and give us a certain knowledge, there remains much to be learned about the origin of the conductance oscillation.

In this Rapid Communication, we carry out a first-principles study on the electron conduction properties of single-row sodium nanowires for a detailed interpretation of the conductance oscillation. We observed that the nanowires manifest the even-odd oscillation of conductance and the channel charge-density distribution forms bunches of high electron density along the nanowire. Moreover, the length of the bunches is equal to the period of the oscillation, 2$d$, where $d$ ($d=7.0$~a.u.) is the nearest-neighbor interatomic distance in the sodium bulk. The relation between the conductance oscillation and the emergence of the bunches is addressed by using simplified models.


Our first-principles calculation code is based on the real-space finite-difference method\cite{cheliko1,cheliko2,ono}. Since this approach does not need any basis-function sets, unlike the plane-wave approach or linear-combination-atomic-orbital (LCAO) method, one can eliminate drawbacks due to basis sets, e.g., a strict treatment of the systems with the combination of periodic and nonperiodic boundaries is computationally difficult in the plane-wave approach and the completeness of the basis set is always a concern in the LCAO method. These advantages allow us to calculate the conduction properties of nanostructures suspended between semi-infinite electrodes with a high degree of accuracy.

\begin{figure}[ht]
\begin{center}
\includegraphics[width=245pt,height=125pt]{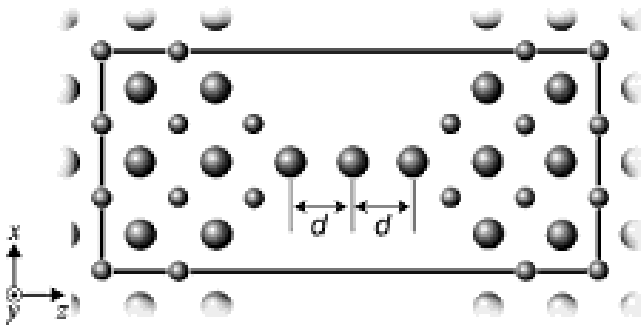}
\caption{Schematic view of single-row sodium nanowire ($N_{atom}=3$) suspended between semi-infinite Na(001) crystalline electrodes. The rectangle represents the supercell employed to determine the Kohn--Sham effective potential and the scattering region to calculate the conduction property.}
\label{egami:fig1}
\end{center}
\end{figure}

We first examine the conduction properties of the single-row sodium nanowires sandwiched between semi-infinite Na(001) crystalline electrodes. Figure~\ref{egami:fig1} shows the computational model in which the case of $N_{atom}=3$ is depicted as an example, where $N_{atom}$ is the number of atoms in the nanowire. The scattering region consists of the single-row sodium wire, the square bases made of sodium atoms, and four sodium (001) plane of electrodes. The nanowire is connected to the bases on both ends and all of them are suspended between the electrodes. The distance between the electrode surface and the basis, as well as that between the basis and the edge atom of the nanowire, is $d/\sqrt{3}$ and the distance between adjacent atoms in the nanowire is $d$. We treat a local pseudopotential of a sodium $3s$ electron for the nucleus constructed using the algorithm by Bachelet, Hamann, and Schl\"{u}ter\cite{bhs}. Exchange-correlation effects are treated by the local density approximation\cite{lda} within the framework of the density functional theory\cite{hohen}. The central finite-difference formula is employed for the derivative arising from the kinetic-energy operator in the Kohn--Sham equation\cite{kohn}. The grid spacing is taken to be 0.81~a.u., which corresponds to a cutoff energy of 15~Ry in the plane-wave method. The Kohn--Sham effective potential is determined self-consistently using the conventional periodic supercell which is represented by the rectangle in Fig.~\ref{egami:fig1}. In the calculation of the global wave functions for infinitely extended states continuing from one electrode side to the other, the overbridging boundary-matching method\cite{obm} is adopted with the supercell imposed periodic boundary conditions in the $x$ and $y$ directions and the nonperiodic boundary condition in the $z$ direction. The electron transmission of the nanowire at the zero bias limit is calculated by the Landauer formula\cite{landauer}.

\begin{figure}[ht]
\begin{center}
\includegraphics[width=210pt,height=145pt]{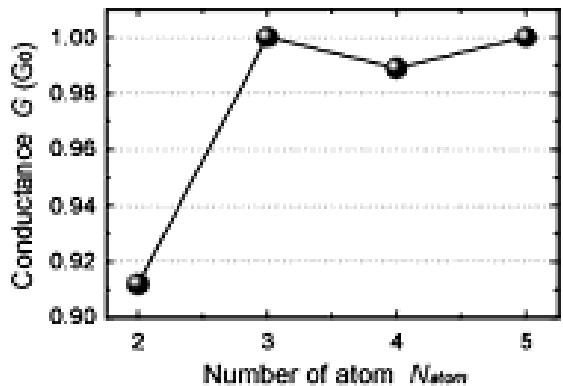}
\caption{Conductances of single-row sodium nanowires.}
\label{egami:fig2}
\end{center}
\end{figure}

\begin{figure*}
\begin{center}
\includegraphics[width=450pt,height=145pt]{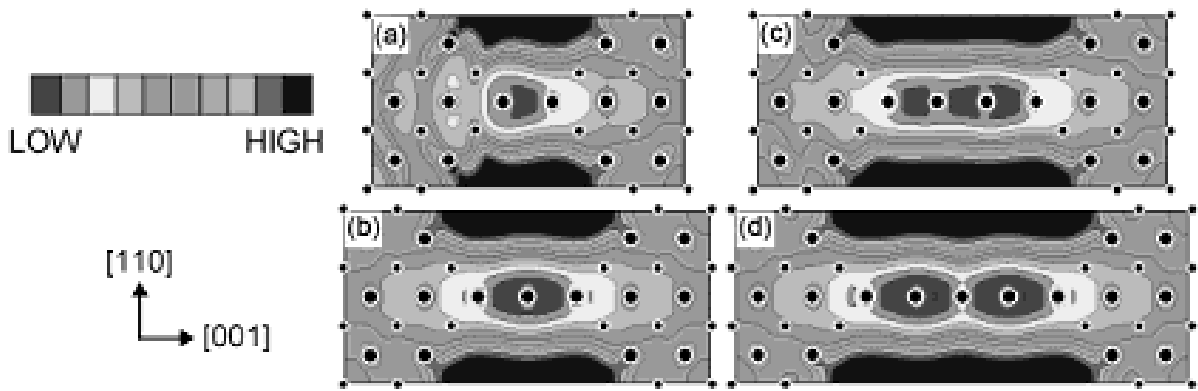}
\caption{(color online). Channel charge-density distributions of incident electrons with the Fermi energy from left-hand electrode for (a) $N_{atom}=2$, (b) $N_{atom}=3$, (c) $N_{atom}=4$ and (d) $N_{atom}=5$. Each contour represents twice or half the density of the adjacent contour lines. The lowest contour represents $4.84 \times 10^{-9}$ electron/$\mbox{bohr}^3$/eV.}
\label{egami:fig3}
\end{center}
\end{figure*}

Figures~\ref{egami:fig2} and \ref{egami:fig3} depict the conductances and the channel charge-density distributions of incident electrons with the Fermi energy from the left-hand electrode, respectively. The planes shown in Fig.~\ref{egami:fig3} are perpendicular to the [100] direction and include the nanowire axis. The conductances of nanowires are $\sim~1~{\rm G_0}$, which are in good agreement with other theoretical and experimental results\cite{krans1,krans2,sim,lang1,lang2,lang3,kobayashi1,kobayashi2,tsukamoto,lee,egami,petr}. In Fig.~\ref{egami:fig2}, one can evidently recognize that the conductances of the odd-number nanowires ($N_{atom}=3$ and 5) are higher than those of the even-number nanowires ($N_{atom}=2$ and 4); the conductance oscillation with a period of two atom-lengths is consistent with that reported in the other theoretical studies\cite{sim,lang1,lang2,lang3,kobayashi1,kobayashi2,tsukamoto,lee,egami,petr} and experiment\cite{smit}. Additionally, in the case of the odd-number nanowires, one observes bunches of high electron density with a length of $2d$ [Figs.~3(b) and \ref{egami:fig3}(d)] which corresponds to the period of the even-odd oscillation, and the density distributions of the incident electrons are almost symmetric with respect to the (001) plane including the center atom of the nanowire.

Next, we will explore the relationship between the even-odd oscillation and the bunches of high electron density. For simplicity, we first consider the one-dimensional problem of the penetration of a square potential barrier as shown in Fig.~4(a), where the nanowire is imitated by the potential barrier. An incident electron with an energy $E$ has a wave vector of $k=\sqrt{2(E-V_0)}$ in the barrier region and its transmission $T$ is determined by wave-function matching\cite{wfmf}
\begin{equation}
T=\frac{4 \kappa^2 k^2}{4 \kappa^2 k^2+(\kappa^2 -k^2)^2\sin^2(kL)},
\label{eqn1}
\end{equation}
where $V_0$ is the height of the potential barrier, $L$ is the barrier length and $\kappa=\sqrt{2E}$. When $L$ is an integral multiple of $\pi/k$, the transmission shows unity. The electron density forms bunches with a length of $\pi/k$ so that it becomes almost symmetric with respect to the center of the nanowire. On the other hand, when $L$ is half-integral multiple of $\pi/k$, the transmission exhibits minimum and the density distribution becomes asymmetric due to the fractured bunches. More realistically, we then consider a three-dimensional prismatic jellium wire with side lengths of $W$ connected to semi-infinite electrodes [Fig.~4(b)]. The potentials for electrons are defined as zero inside the nanowire and electrodes, whereas that in the vacuum regions is infinitely high. It is well known that the lowest band energy of the infinite prismatic jellium wire with $W$ in the $x$ and $y$ directions is $E=\frac{1}{2}k_z^2+(\pi/W)^2$, where $k_z$ is the $z$ component of the wave vector. When the wire is connected to the electrodes, the incident electrons with the energy $E$ incoming from the electrode have a wave vector of $k_z = \sqrt{2[E-(\pi/W)^2]}$ in the nanowire region and the transmission $T$ is determined by
\begin{equation}
T=\frac{k_z^2(\beta+\beta^{\ast})^2}{k_z^2(\beta+\beta^{\ast})^2+|k_z^2-\beta^2|^2\sin^2[{k_z}(L-\Delta L)]},
\label{eqn2}
\end{equation}
where $\beta=-{\it i}\frac{d}{dz} \Psi^{in}(0)$ with $\Psi^{in}(z)$ being the channel wave function of incident electrons from the left-hand electrode and $\Delta L=\frac{1}{2 \pi i} \log \left[ \frac{(k_z+\beta^{\ast})(k_z-\beta)}{(k_z+\beta)(k_z-\beta^{\ast})} \right]$. Note that $\Psi^{in}(z)$ includes left-decreasing evanescent waves as well as right-propagating waves.
Figures~\ref{egami:fig4} and \ref{egami:fig5} show the transmission as a function of the nanowire length and channel charge-density distributions of the incident electrons from the left-hand electrode, respectively. The period of the oscillation and the length of the bunches do not depend on the energy of incident electrons, when they are measured in units of $\pi/k_z$. Furthermore, when $L$ is an integral multiple of $\pi/k_z$, the conductance is $1~{\rm G_0}$ and the channel density distribution is symmetric with respect to the center of the nanowire. On the other hand, when $L$ is a half-integral multiple of $\pi/k_z$ (Fig.~\ref{egami:fig5}), the conductance exhibits minimum and the distribution becomes asymmetric.
\begin{figure}[ht]
\begin{center}
\includegraphics[width=270pt,height=205pt]{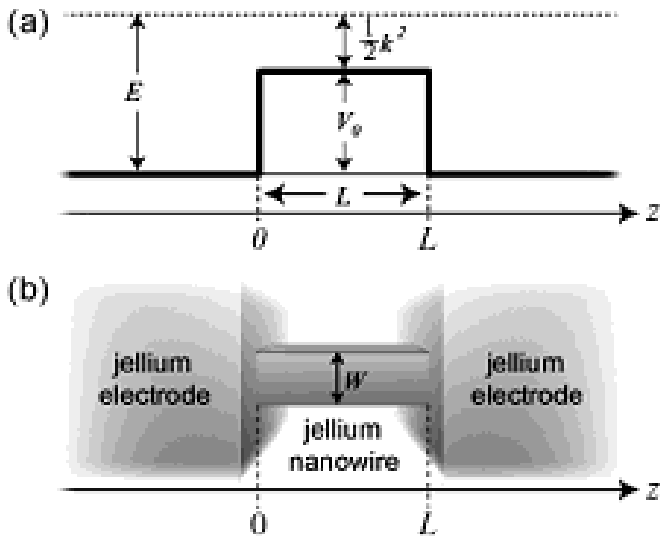}
\caption{Schematic view of (a) potential barrier in one-dimensional system and (b) prismatic jellium wire connected to semi-infinite jellium electrodes.}
\label{egami:fig6}
\end{center}
\end{figure}

\begin{figure}[ht]
\begin{center}
\includegraphics[width=240pt,height=175pt]{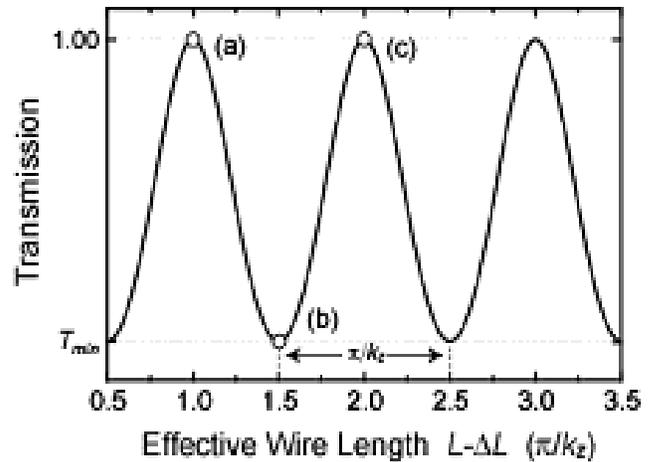}
\caption{Conductance of prismatic jellium nanowire as function of wire length $L$. Here, $T_{min}$ is $[1+|k_z^2-\beta^2|^2/k_z^2(\beta+\beta^{\ast})^2]^{-1}$. (a), (b) and (c) are the points where the channel charge-density distributions are shown in Fig.~\ref{egami:fig5}.}
\label{egami:fig4}
\end{center}
\end{figure}

\begin{figure*}
\begin{center}
\includegraphics[width=505pt,height=150pt]{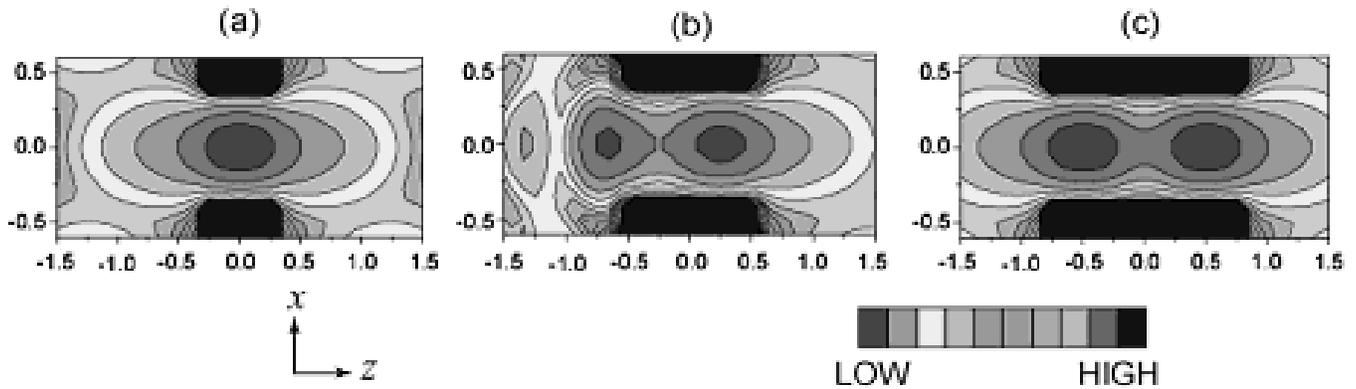}
\caption{(color online). Channel charge-density distributions of incident electrons from left-hand electrode at (a), (b) and (c) in Fig.~\ref{egami:fig4}. Each contour represents twice or half the density of the adjacent contour lines.}
\label{egami:fig5}
\end{center}
\end{figure*}

Now, we return to the case of the atomistic sodium nanowires. Since the energy band of the infinite single-row sodium wire with the interatomic distance of $d$ crosses the Fermi energy at the wave vector $k_z$ of $\pi/2d$\cite{ono2}, the period of the conductance oscillation and the length of the bunches become $2d$, which justifies the above-mentioned explanation.

Smit {\it et al.}\cite{smit} and Lee {\it et al.}\cite{lee} discussed the even-odd oscillation behavior using one-dimensional nanowires with finite and infinite lengths, and however they did not mention the emergence of the bunches of high electron density with relation to the conductance oscillation.


In summary, we have studied the relation between the even-odd oscillation of conductance and the bunches of high electron density by calculating the conduction properties of atomistic sodium nanowires and jellium ones. The conductance oscillation can be understood in terms of the characteristics of the infinite wire and the conventional wave-function matching: when the nanowire whose length is an integral multiple of $\pi/k_z$ is connected to the electrodes, the conductance is maximum and bunches with a length of $\pi/k_z$ come about so that the channel density distribution becomes symmetric. On the other hand, in the case where the nanowire length is a half-integral multiple of $\pi/k_z$, the conductance is minimum and the distribution becomes asymmetric. The work in progress is the investigation of nanowires consisting of multivalent atoms.

This research was supported by a Grant-in-Aid for the 21st Century COE ``Center for Atomistic Fabrication Technology'' and also by a Grant-in-Aid for Scientific Research (C) (Grant No. 16605006) from the Ministry of Education, Culture, Sports, Science and Technology. The numerical calculation was carried out with the computer facilities at the Institute for Solid State Physics at the University of Tokyo, and the Information Synergy Center at Tohoku University.


\end{document}